\documentclass[aps,prb,twocolumn,groupedaddress,showkeys,showpacs]{revtex4}

\newcommand{\op}[1]{%
    \fontdimen12\textfont3=2pt\fontdimen12\scriptfont3=1.4pt%
    \!\null\mathop{\vphantom{#1}\smash{#1}}\limits_{\sim}\null\!}
\newcommand{\xref}[1]{\protect\ref{#1}}

\newcommand{\fmref}[1]{(\protect\ref{#1})}
\def\bra#1{\langle \, {#1} \, | \,}
\def\ket#1{\, | \, {#1} \, \rangle}
\newcommand{\braket}[2]{\langle \, {#1} \, | \, {#2} \, \rangle}
\newcommand{\vek}[1]{{\!\vec{\,#1}}}
\newtheorem{k-rule}{k-rule}

\usepackage{multirow,amssymb,amsbsy,amsmath}
\usepackage{epsfig}

\begin{document}

\title{Quantum numbers for relative ground states of
  antiferromagnetic Heisenberg spin rings}

\author{Klaus B\"arwinkel}
\email{Klaus.Baerwinkel@uos.de}
\affiliation{Universit\"at Osnabr\"uck, Fachbereich Physik,
D-49069 Osnabr\"uck, Germany}

\author{Peter Hage}
\email{phage@uos.de}
\affiliation{Universit\"at Osnabr\"uck, Fachbereich Physik,
D-49069 Osnabr\"uck, Germany}

\author{Heinz-J\"urgen Schmidt}
\email{hschmidt@uos.de}
\affiliation{Universit\"at Osnabr\"uck, Fachbereich Physik,
D-49069 Osnabr\"uck, Germany}

\author{J\"urgen Schnack}
\email{jschnack@uos.de}
\affiliation{Universit\"at Osnabr\"uck, Fachbereich Physik,
D-49069 Osnabr\"uck, Germany}

\date{\today}

\begin{abstract}
We suggest a general rule for the shift quantum numbers $k$
of the relative ground states of antiferromagnetic Heisenberg
spin rings. This rule generalizes well-known results of Marshall,
Peierls, Lieb, Schultz, and Mattis for even rings. Our rule is
confirmed by numerical investigations and rigorous proofs for
special cases, including systems with a Haldane gap for
$N\rightarrow\infty$. Implications for the total spin quantum
number $S$ of relative ground states are discussed as well as
generalizations to the $XXZ$ model.
\end{abstract}

\maketitle

\section{Introduction}

Rigorous results on spin systems like the Marshall-Peierls sign
rule \cite{Mar:PRS55} and the famous theorems of Lieb, Schultz,
and Mattis \cite{LSM:AP61,LiM:JMP62} have sharpened our
understanding of magnetic phenomena. They as well serve as
theoretical input for quantum computing with spin
systems.\cite{WaZ:PLA02,Wang:PRA02A,Wang:PRA02B}

Exact diagonalization methods allow to evaluate eigenvalues and
eigenvectors of $\op{H}$ for small spin rings of various numbers
$N$ of spin sites and spin quantum numbers $s$ where the
interaction is given by antiferromagnetic nearest neighbor
exchange
\cite{BoF:PR64,BoJ:PRB83,FLM:PRB91,Kar94,BSS:JMMM00:B,Schnack:PRB00}.
One quantity of interest is the shift quantum number $k=0,\dots
N-1$ associated with the cyclic shift symmetry of the rings.
Using the sign rule of Marshall and Peierls \cite{Mar:PRS55} or
equivalently the theorems of Lieb, Schultz, and Mattis
\cite{LSM:AP61,LiM:JMP62} one can explain the shift quantum
numbers for the relative ground states in subspaces ${\mathcal
H}(M)$ of total magnetic quantum number $M$ for rings with
even $N$. In the case of $s=1/2$ one knows the shift quantum
numbers of the total ground states for all $N$ via the Bethe
ansatz.\cite{Kar94}

The sign rule of Marshall and Peierls as well as the theorems of
Lieb, Schultz, and Mattis only apply to bipartite rings,
i.~e.~rings with even $N$.  Nevertheless, even for frustrated
rings with odd $N$ astonishing regularities are numerically
verified. This creates the need for a deeper insight or -- at
best -- an analytic proof for the simple k-rule~\ref{c-1}
which comprises all these results. Unifying the picture for even
and odd $N$, we find for the ground state without
exception:\cite{BSS:JMMM00:B,Schnack:PRB00}
\begin{enumerate}
\item The ground state belongs to the subspace ${\mathcal H}(S)$
with the smallest possible total spin quantum number $S$;
this is either $S=0$ for $N\!\cdot\!s$ integer, then the total magnetic quantum
number $M$ is also zero, or $S=1/2$ for $N\!\cdot\!s$ half integer, then
$M=\pm 1/2$.
\item If $N\!\cdot\!s$ is integer, then the ground state is non-degenerate.
\item If $N\!\cdot\!s$ is half integer, then the ground state is fourfold degenerate.
\item If $s$ is integer or $N\!\cdot\!s$ even, then the shift
quantum number is $k=0$.
\item If $s$ is half integer and $N\!\cdot\!s$ odd, then the shift
quantum number turns out to be $k=N/2$.
\item If $N\!\cdot\!s$ is half integer, then
$k=\lfloor(N+1)/4\rfloor$ and $k=N-\lfloor(N+1)/4\rfloor$ is
found.  $\lfloor(N+1)/4\rfloor$ symbolizes the greatest integer
less than or equal to $(N+1)/4$.
\end{enumerate}

\begin{table}[ht!]
\begin{center}
\begin{tabular}{|c|c||l|l|l|l|l|l|l|l|l|l|l|}
\hline
$N$&$s$&\multicolumn{11}{c|}{$a$}\\
&& 1&2&3&4&5&6&7&8&9&10&11\\
\hline\hline
3 & 1/2 & 1,2  &  -  &  -  &  -  &  -  &  -  &  -  &  -  &  -  & - &  -  \\
3 & 1   & 1,2  &\textbf{0},1,2&  0  &  -  &  -  &  -  &  -  &  -  &  -  & - &  -  \\
3 & 3/2 & 1,2  &\textbf{0},1,2&0,\textbf{1},\textbf{2}& 1,2 &  -  &  -  &  -  &  -  &  -  & - &  -  \\
3 & 2   & 1,2  &\textbf{0},1,2&0,\textbf{1},\textbf{2}&\textbf{0},1,2&\textbf{0},1,2&  0  &  -  &  -  &  -  & - &  -  \\
\hline
5 & 1/2 & 2,3  & 1,4 &  -  &  -  &  -  &  -  &  -  &  -  &  -  & - &  -  \\
5 & 1   & 2,3  & 1,4 & 1,4 & 2,3 &  0  &  -  &  -  &  -  &  -  & - &  -  \\
5 & 3/2 & 2,3  & 1,4 & 1,4 & 2,3 &  0  & 2,3 & 1,4 &  -  &  -  & - &  -  \\
5 & 2   & 2,3  & 1,4 & 1,4 & 2,3 &  0  & 2,3 & 1,4 & 1,4 & 2,3 & 0 &  -  \\
\hline
7 & 1/2 & 3,4  & 1,6 & 2,5 &  -  &  -  &  -  &  -  &  -  &  -  & - &  -  \\
7 & 1   & 3,4  & 1,6 & 2,5 & 2,5 & 1,6 & 3,4 &  0  &  -  &  -  & - &  -  \\
7 & 3/2 & 3,4  & 1,6 & 2,5 & 2,5 & 1,6 & 3,4 &  0  & 3,4 & 1,6 & 2,5 &  -  \\
\hline
9 & 1/2 & 4,5  & 1,8 & 3,6 & 2,7 &  -  &  -  &  -  &  -  &  -  & - &  -  \\
9 & 1   & 4,5  & 1,8 & 3,6 & 2,7 & 2,7 & 3,6 & 1,8 & 4,5 &  0  & - &  -  \\
\hline
11& 1/2 & 5,6  & 1,10& 4,7 & 2,9 & 3,8 &  -  &  -  &  -  &  -  & - &  -  \\
11& 1   & 5,6  & 1,10& 4,7 & 2,9 & 3,8 & 3,8 & 2,9 & 4,7 & 1,10&5,6&  0  \\
\hline
\hline
\end{tabular}
\vspace*{5mm}
\end{center}
\caption[]{Numerically verified shift quantum numbers for
  selected $N$ and $s$ in subspaces ${\mathcal H}(M)$. Instead
  of $M$ the quantity $a=Ns-M$ is used. The shift quantum number
  for the magnon vacuum $a=0$ is always $k=0$. The shift quantum
  numbers are invariant under $a\leftrightarrow
  Ns-a$ and hence only displayed for $a=1,2,\ldots,\lfloor Ns\rfloor $.
  Extraordinary shift quantum numbers given in
  bold do not comply with Eq.~\ref{E-1-1}.}\label{T-1}
\end{table}
In this article we will extend the knowledge about shift quantum
numbers for the relative ground states in subspaces ${\mathcal
H}(M)$ for odd rings. Table \xref{T-1} shows a small selection
of shift quantum numbers for some $N$ and $s$. The dependence of
$k$ on $N$ and $a$ can -- for even as well as for odd $N$ -- be
generalized as given by the following
\begin{k-rule}
\label{c-1}
\begin{eqnarray}
\label{E-1-1}
\text{If}\;N\ne 3
&\text{ then }&\;
k
\equiv
\pm
a
\lceil
\frac{N}{2}
\rceil
\mod N
\ .
\end{eqnarray}
Moreover the degeneracy of the relative ground state is minimal.
\end{k-rule}
Here $\lceil N/2\rceil$ denotes the smallest integer greater
than or equal to $N/2$. ``Minimal degeneracy" means that the
relative ground state in ${\mathcal H}(M)$ is twofold degenerate
if there are two different shift quantum numbers and
non-degenerate if $k=0$ mod $N$ or $k=N/2$ mod $N$, the latter
for even $N$.

It is noteworthy that the shift quantum numbers do not depend on
$s$. For $N=3$ and $3s-2\ge |M| \ge 1$ we find besides the
ordinary shift quantum numbers given by \fmref{E-1-1}
extraordinary shift quantum numbers, which supplement the
ordinary ones to the complete set $\{k\}=\{0,1,2\}$. This means
an additional degeneracy of the respective relative ground
state, which is caused by the high symmetry of the Heisenberg
triangle.

For even $N$ the k-rule \fmref{E-1-1} results in an
alternating $k$-sequence $0, N/2, 0, N/2, \dots$ on descending
from the magnon vacuum with $M=Ns$, i.~e. $a=0$, which
immediately implies that the ground state in ${\mathcal H}(M)$
has the total spin quantum number $S=|M|$, compare
Refs.~\onlinecite{Mar:PRS55,LSM:AP61,LiM:JMP62}.

For odd $N$ the regularity following from \fmref{E-1-1}
will be illustrated by an example: Let $N=11$ and $s$ be
sufficiently large.  Then the $k$-sequence reads $0,\pm 6,\pm
1,\pm 7,\pm 2,\pm 8,\pm 3,\pm 9, \pm 4,\pm 10,\pm 5,0,\ldots$,
where all shift quantum numbers are understood mod $11$.  The
sequence is periodic with period $11$ and repeats itself after
$5$ steps in reverse order. In the first $5$ steps each possible
$k$-value is assumed exactly once. Since $\pm 8 =\mp 3 \mod 11$,
the shift quantum numbers for $a=5$ and $a=6$ 
are the same, likewise for $a=16$ and $a=17$ and so on.

The last finding can be easily generalized: For odd $N$ the $k$
quantum numbers are the same in adjacent subspaces ${\mathcal
H}(M=Ns-a)$ and ${\mathcal H}(M^\prime=Ns-(a+1))$ iff $N$
divides $(2a+1)$. In such cases one cannot conclude that the
ground state in ${\mathcal H}(M)$ has the total spin quantum
number $S=|M|$, nevertheless, in all other cases including the
total ground state one can, see section \ref{sec-3}.

The k-rule \ref{c-1} is founded in a mathematically rigorous
way for $N$ even,\cite{Mar:PRS55,LSM:AP61,LiM:JMP62} $N=3$
(including extraordinary $k$ numbers, see section \ref{sec-4-3}),
$a=0$ (trivial), $a=1$ (cf.~section \ref{sec-4-1}), $a=2$ (but
only in a weakened version, cf.~section \ref{sec-4-4}).  For the
ground state with $N$ odd, $s=1/2$ the k-rule follows from the
Bethe ansatz, cf. section \ref{sec-4-2}. An asymptotic proof for
large enough $N$ is provided in section \ref{sec-4-5} for
systems with an asymptotically finite excitation gap (Haldane
systems). The k-rule also holds for the exactly solvable
$XY$-model with $s=1/2$, cf.~section \ref{sec-6}.  Apart from
these findings a rigorous proof of the k-rule still remains a
challenge.

\section{Heisenberg model}
\label{sec-2}

The Hamilton operator of the Heisenberg model with
antiferromagnetic, isotropic nearest neighbor interaction between
spins of equal spin quantum number $s$ is given by
\begin{eqnarray}
\label{E-2-1}
\op{H}
&\equiv&
2\,
\sum_{i=1}^N\;
\op{\vec{s}}_i \cdot \op{\vec{s}}_{i+1}
\ ,\quad N+1\equiv 1
\ .
\end{eqnarray}
$\op{H}$ is invariant under cyclic shifts generated by the shift
operator $\op{T}$. $\op{T}$ is defined by its action on the
product basis $\ket{\vec{m}}$ \fmref{E-2-2}
\begin{eqnarray}
\label{E-2-3}
\op{T}\,
\ket{m_1, \dots, m_N}
\equiv
\ket{m_N, m_1, \dots, m_{N-1}}
\ ,
\end{eqnarray}
where the product basis is constructed from single-particle
eigenstates of all $\op{s}^3_i$
\begin{eqnarray}
\label{E-2-2}
\op{s}^3_i\,
\ket{m_1, \dots, m_N}
=
m_i\,
\ket{m_1, \dots, m_N}
\ .
\end{eqnarray}
The shift quantum number $k=0,\dots, N-1$ modulo $N$ labels the
eigenvalues of $\op{T}$ which are the $N$-th roots of unity
\begin{eqnarray}
\label{E-2-4}
z
=
\exp\left\{
-i \frac{2\pi k}{N}
\right\}
\ .
\end{eqnarray}
Altogether $\op{H}$, $\op{T}$, the square $\op{\vec{S}}^2$, and
the three-component $\op{S}^3$ of the total spin are four
commuting operators. The subspaces of states with the
quantum numbers $M,S,k$ will be denoted by ${\mathcal
H}_N(M,S,k)$.

The Hamilton operator \fmref{E-2-1} can be cast in the form
\begin{eqnarray}
\label{E-2-5}
\op{H}
&=&
\op{\Delta} + \op{G} + \op{G}^\dagger
\ ,
\end{eqnarray}
where we introduced
\begin{eqnarray}
\label{E-2-7}
\op{\Delta}
&\equiv&
2\,
\sum_{i=1}^N\;
\op{s}_i^3 \op{s}_{i+1}^3
\ ,
\end{eqnarray}
and the ``generation operator"
\begin{eqnarray}
\label{E-2-6}
\op{G}
&\equiv&
\sum_{i=1}^N\;
\op{s}_i^- \op{s}_{i+1}^+
\end{eqnarray}
together with its adjoint $\op{G}^\dagger$.

It follows that $\op{H}$ is represented by a real matrix with
respect to the product basis.  Hence if an eigenvector of this
matrix has the shift quantum number $k$, its complex conjugate
will be again an eigenvector with the same eigenvalue but with
shift quantum number $-k$ mod $N$. Simultaneous eigenvectors of
$\op{H}$ and $\op{T}$ can be chosen as real in the product basis
only if $k=0$ or $k=N/2$.

We define a unitary ``Bloch" operator $\op{U}$ for spin rings,
compare Refs.~\onlinecite{LSM:AP61,AfL:LMP86},
\begin{eqnarray}
\label{E-4-1}
\op{U}
&\equiv&
\exp\left\{
\frac{2 \pi i}{N}\, \sum_{j=1}^N\, j\, (s-\op{s}_j^3)
\right\}
\ ,
\end{eqnarray}
which is diagonal in the product basis \fmref{E-2-2}.

We then have
\begin{eqnarray}
\label{E-4-2-B}
\op{T}\op{U}\op{T}^{\dagger}\op{U}^{\dagger}
&=&
\exp\left\{
-\frac{2 \pi i}{N}\, \sum_{j=1}^N\, (s-\op{s}_{j}^3)
\right\}
\\
\label{E-4-2-C}
&=&
\exp\left\{
-\frac{2 \pi i}{N}\, a
\right\}
\ ,
\end{eqnarray}
where the last line \fmref{E-4-2-C} holds in subspaces
${\mathcal H}(M=Ns-a)$.  Consequently, $\op{U}$ is a shift
operator in $k$-space and shifts the quantum number $k$ of a
state $\ket{\phi}\in {\mathcal H}(M)$ by $a$:
\begin{eqnarray}
\label{E-4-2-D}
\mbox{If }
\op{T}\ket{\phi}
&=&
\exp\left\{
-\frac{2 \pi i}{N}\, k
\right\}
\ket{\phi}
\\
\mbox{ then }
\op{T}\op{U}\ket{\phi}
&=&
\exp\left\{
-\frac{2 \pi i}{N}\, (k+a)
\right\}
\op{U}\ket{\phi}
\nonumber
\ .
\end{eqnarray}
We also observe that
\begin{eqnarray}
\label{E-4-2-E}
\op{U}\op{G}\op{U}^\dagger
&=&
\exp\left\{
-\frac{2 \pi i}{N}
\right\}
\op{G}
\ ,
\end{eqnarray}
and define the unitary ``Bloch" transform of the Hamilton
operator
\begin{eqnarray}
\label{E-4-3}
\widehat{\op{H}}(\ell)
\equiv
\op{U}^\ell \op{H} (\op{U}^\dagger)^\ell
&=&
\op{\Delta}
+ \cos\left( \frac{2 \pi \ell}{N} \right)
\left\{ \op{G} + \op{G}^\dagger\right\}
\\
&&
- i \sin\left( \frac{2 \pi \ell}{N} \right)
\left\{ \op{G} - \op{G}^\dagger\right\}
\nonumber
\ .
\end{eqnarray}
If we choose $\ell=\ell(N)=\pm\lceil N/2\rceil$, then
$\cos\left( \frac{2 \pi \ell}{N} \right)$ is as close to $-1$ as
possible. We will use the short-hand notation
$\op{H}_{\text{B}}\equiv\widehat{\op{H}}(\lceil N/2\rceil)$ and
equation \fmref{E-4-2-D} then yields a relation between the
eigenstates of $\op{H}_{\text{B}}$ and $\op{H}$: If any
eigenstate $\ket{\Psi_B}$ of $\op{H}_{\text{B}}$ has the shift
quantum number $k_B$ then the corresponding eigenstate of the
original Hamiltonian has the shift quantum number $k=k_B - a
\lceil N/2\rceil$.

Consequently the k-rule \ref{c-1} is equivalent to
\begin{k-rule}
\label{c-2}
For $N\ne 3$ the relative ground states of $\op{H}_{\text{B}}$
have the shift quantum numbers 
\begin{eqnarray}\label{e-c-2}
k =
\left\{
\begin{array}{r@{\quad:\quad}l}
0 \mod N & N\ \text{even}
\\
0, a \mod N & N\ \text{odd}
\end{array}
\right.
\ .
\end{eqnarray}
Their degeneracy is minimal.
\end{k-rule}
For later use we also define a ``Frobenius-Perron" Hamiltonian
as
\begin{eqnarray}
\label{E-4-4}
\op{H}_{\text{FP}}(x)
&=&
\op{\Delta} +
x
\;
\left\{ \op{G} + \op{G}^\dagger\right\}
\ ,
\end{eqnarray}
where $x$ is an arbitrary real number. For negative $x$ the
operator \fmref{E-4-4} satisfies the conditions of the theorem
of Frobenius and Perron\cite{Lan:69} with respect to the
product basis. We will utilize the following version of this
theorem, adapted to the needs of physicists:

Let a symmetric matrix ${\mathcal A}$ have off-diagonal elements
$\leq 0$. Moreover, let ${\mathcal A}$ be \emph{irreducible},
which means that every matrix element of ${\mathcal A}^n$ is
non-zero for sufficiently high powers $n$ of ${\mathcal A}$. Then
${\mathcal A}$ has a non-degenerate ground state with positive
components.  

Thus, in our case and for odd $N$ the ground state of
$\op{H}_{\text{FP}}(x)$ will have the shift quantum number
$k=0$.

The Bloch transform for even $N$ results in a pure
Frobenius-Perron Hamiltonian,
i.~e. $\op{H}_{\text{B}}=\op{H}_{\text{FP}}(-1)$, whereas for
odd $N$ one obtains
\begin{eqnarray}
\label{E-4-7}
\op{H}_{\text{B}}
=
\op{H}_{\text{FP}}(-\cos\left( \frac{\pi}{N} \right))
- i \sin\left( \frac{\pi}{N} \right)
\left\{ \op{G} - \op{G}^\dagger\right\}
\ .
\end{eqnarray}

\section{Consequences of the k-rule}
\label{sec-3}

In the following we only consider the new case of odd $N$ since
the respective relations for even $N$ are already known for a
long time.\cite{Mar:PRS55,LSM:AP61,LiM:JMP62}

Subspaces ${\mathcal H}(M)$ and ${\mathcal H}(M^\prime)$ are
named ``adjacent" if $M^\prime=M-1$ or, equivalently,
$a^\prime=a+1$. The ordinary $k$-numbers for the respective
relative ground states are $k=\pm a \lceil N/2\rceil\mod N$
and $k^\prime=\pm (a+1) \lceil N/2\rceil\mod N$.  As mentioned
above these quantum numbers are different unless $N$ divides
$2a+1$.

Relative ground states can be chosen to be eigenstates of
$\op{\vec{S}}^2$. As we are going to show, the k-rule helps to
understand that the total spin quantum number $S$ of a relative
ground state in ${\mathcal H}(M\ge 0)$ is $S=M$ not only for
even $N$ but also for odd $N$.

Let us consider $M^\prime=Ns-(a+1)\ge 0$ and let
$\ket{\phi_k(a+1)}$ be a ground state in ${\mathcal
H}(M^\prime)$. If this state vanishes on applying the total
ladder operator $\op{S}^+=\sum_i \op{s}_i^+$, it is an
eigenstate of $\op{\vec{S}}^2$ with $S=M^\prime=Ns-(a+1)$.

The question is now whether $\op{S}^+\ket{\phi_k(a+1)}\ne 0$ is
possible? If so, the resulting state would be an eigenstate of
the shift operator $\op{T}$ with the same $k$-number,
i.~e. $k=\pm(a+1)\lceil N/2\rceil$. But on the other hand the
resulting state is also a ground state in ${\mathcal
H}(M=Ns-a)$, because all the energy eigenvalues in ${\mathcal
H}(M=Ns-a)$ are inherited by ${\mathcal
H}(M^\prime=Ns-(a+1))$. Then, the k-rule applies, but now for
$a$ instead of $(a+1)$, which produces a contradiction unless
for those cases where $N$ divides $(2a+1)$. In the latter cases
one cannot exclude that the relative ground state energies
$E_{\text{min}}(M)$ and $E_{\text{min}}(M^\prime)$ are the
same. 

We thus derive an $S$-rule from the k-rule for odd $N$:
\begin{itemize}
\item If $N$ does not divide $(2a+1)$, then any relative ground
  state in ${\mathcal H}(M=Ns-(a+1))$ has the total spin quantum
  number $S=|M|$. In accordance the minimal energies fulfill
  $E_{\text{min}}(M=S)<E_{\text{min}}(M=S+1)$.
\item For the absolute ground state with $a+1=Ns$ or
  $a+1=Ns-1/2$, $N$ does never divide $(2a+1)$. The k-rule
  therefore yields, that the total spin of the absolute ground
  state is $S=0$ for $Ns$ integer and $S=1/2$ for $Ns$ half
  integer.
\end{itemize}

As an example we would like to discuss the case of $N=5$ and
$s=1$, compare Table~\ref{T-1}. The magnon vacuum $a=0$ has the
total magnetic quantum number $M=Ns=5$, $k=0$, and $S=Ns=5$. The
adjacent subspace with $a=1$ has $M=4$ and $k=2,3$, therefore,
the ground state in this subspace must have $S=4$. If the ground
state had $S=5$ it would already appear in the subspace
``above". The next subspace belongs to $a=2$, i.~e. $M=3$. It
again has a different $k$, thus $S=3$. While going to the next
subspace ${\mathcal H}(M)$ the $k$-number does not
change. Therefore, we cannot use our argument. We only know that
the minimal energy in this subspace is smaller than or equal to
that of the previous subspace. Going further down in $M$ the
$k$-values of adjacent subspaces are again different, thus
$S=|M|$ and $E_{\text{min}}(M=S)<E_{\text{min}}(M=S+1)$.

\section{Proofs for special cases}
\label{sec-4}

\subsection{The case $a=1$}
\label{sec-4-1}

The eigenvalues of the Hamiltonian in the subspace with $a=1$
are well-known:
\begin{eqnarray}\label{5.1.1}
E_k
&=&
2 N s^2 - 4 s + 4 s  \cos \frac{2\pi k}{N}
\ ,
\\
&&
k=0,1,\ldots,N-1
\ ,
\nonumber
\end{eqnarray}
where $k$ is the corresponding shift quantum number. Obviously,
the relative ground state is obtained for $k=\frac{N}{2}$ for
even $N$ and $k=\frac{N\pm 1}{2}$ for odd $N$.

\subsection{The ground state of odd $s=1/2$ rings}
\label{sec-4-2}

In this case the ground state belongs to $a=\frac{N-1}{2}$ and
the k-rule \fmref{E-1-1} reads
\begin{eqnarray}\label{5.2.1}
k
=
\pm a^2 \mod N
=
\pm \left(\frac{N-1}{2}\right)^2\mod N
\ .
\end{eqnarray}
It is an immediate consequence of the Bethe ansatz as we will
show. Following the notation of Ref.~\onlinecite{Yosida91},
chapter 9.3, the energy eigenvalues in the subspace with $M=1/2$
may be written as
\begin{eqnarray}\label{5.2.2}
 E=2\epsilon-N/2,
\end{eqnarray}
with
\begin{eqnarray}\label{5.2.3}
 \epsilon=\sum_{i=0}^a (1-\cos f_i)
\end{eqnarray}
and
\begin{eqnarray}\label{5.2.4}
N f_i = 2\pi \lambda_i +\sum_j \varphi_{ij}
\ ,
\end{eqnarray}
where the $\lambda_i$ are natural numbers between $0$ and $N-1$
satisfying $|\lambda_i-\lambda_j|\ge 2$ for $i\neq j$ and the
$\varphi_{ij}$ are the entries of some antisymmetric phase
matrix. Hence the two ground state configurations are
$\vec{\lambda}=(1,3,5,\ldots,N-2)$ and
$\vec{\lambda}^\prime=(2,4,6,\ldots,N-1)=-\vec{\lambda}\mod N$.
According to Ref.~\onlinecite{Yosida91}, p.~137, the shift
quantum number of the ground state will be
\begin{eqnarray}\label{5.2.5}
k= \sum_j \lambda_j = \pm a^2\mod N
\ ,
\end{eqnarray}
in accordance with \fmref{5.2.1}.

\subsection{The case $N=3$}
\label{sec-4-3}

In this subsection we want to prove that the shift quantum
numbers $k$ of relative ground states satisfy the rule
\begin{eqnarray}\label{5.3.0}
k =
\left\{
\begin{array}{r@{\quad:\quad}l}
1, 2 & a=1
\\
 0 & a=3s, s\mbox{ integer}
\\
1, 2 & a=3s-1/2, s\mbox{ half integer}
\\
0, 1, 2 & \mbox{else}
\end{array}
\right.
\ .
\end{eqnarray}
By completing squares the Hamiltonian can be written in the form
\begin{eqnarray}\label{5.3.1}
\op{H}
=
\op{\vec{S}}^2-3 s(s+1)
\end{eqnarray}
and can be diagonalized in terms of Racah $6j$-symbols. The
lowest eigenvalues in ${\mathcal H}(M)$ are those with
$S=M=3s-a$.  In order to determine the shift quantum numbers of
the corresponding eigenvectors we may employ the results in
Ref.~\onlinecite{BSS:JMMM00} on the dimension of the spaces
${\mathcal H}_N(M,S,k)$. Using equations (11) and (12) of
Ref.~\onlinecite{BSS:JMMM00} we obtain after some algebra
\begin{eqnarray}\label{5.3.3}
\mbox{dim}({\mathcal H}_3(M,S&=&M))
=
\\
&&
\left\{
\begin{array}{r@{\quad:\quad}l}
a+1 & 0\le a\le 2s \\
6s-2a+1 & 2s\le a \le \lfloor 3s \rfloor
\end{array}
\right.
\ .
\nonumber
\end{eqnarray}
Now consider $\mbox{dim}({\mathcal H}_3(M,k))$.  The product
basis in ${\mathcal H}_3(M)$ may be grouped into $\nu({a})$
proper cycles of three different states
$\{\ket{\vec{m}},\op{T}\ket{\vec{m}},\op{T}^2\ket{\vec{m}}\}$,
and, if $a = 0\mod 3$, one additional state
$\ket{\lambda,\lambda,\lambda}$ having $k=0$. Each
$3$-dimensional subspace spanned by a cycle contains a basis of
eigenvectors of $\op{T}$ with each shift quantum number
$k=0,1,2$ occuring exactly once, hence
\begin{eqnarray}\label{5.3.5}
\mbox{dim}&&({\mathcal H}_3(M,k))
=
\\
&&
\left\{
\begin{array}{l@{\quad:\quad}l}
\nu({a}) & a\neq 0\mod 3  \\
\nu({a}) & k=1, 2 \mbox{ and }a=0 \mod 3\\
\nu({a})+1 & k=0\mbox{ and } a=0 \mod 3.
\end{array}
\right.
\ .
\nonumber
\end{eqnarray}
Note further that $\op{S}^- : {\mathcal H}(M) \longrightarrow
{\mathcal H}(M-1) $ commutes with $\op{T}$, hence maps
eigenvectors of $\op{T}$ onto eigenvectors with the same shift
quantum number. This leads to
\begin{eqnarray}\label{5.3.6}
\mbox{dim}({\mathcal H}_3(M&,&S=M,k))
=
\\
&&
\left\{
\begin{array}{l@{\quad:\quad}l}
\mu({a})+1 & k=0, a=0\mod 3 \\
\mu({a})-1 & k=0, a=1\mod 3 \\
\mu({a})   & k=0, a=2\mod 3 \\
\mu({a})   & k=1, 2
\end{array}
\right.
\ ,
\nonumber
\end{eqnarray}
with
\begin{eqnarray}\label{5.3.6.B}
\mu({a})
&\equiv&
\left\{
\begin{array}{l@{\quad:\quad}l}
0 & a=0 \\
\nu(a)-\nu(a-1) & a>0
\end{array}
\right.
\ .
\end{eqnarray}
Comparison with \fmref{5.3.3} yields those values of $a$ and $s$
where $\mbox{dim}({\cal H}_3(M,S=M,k))$ vanishes for some $k$,
i.~e.~where not all possible shift quantum numbers occur for the
relative ground states.  Due to \fmref{5.3.6} this happens if
$\mu(a)=0$ or $\mu(a)=1$.

For $a=1$ only the values $k=1, 2$ appear according to
subsection \ref{sec-4-1}, hence $\mu(a)=1$.  If $s$ is integer
and $a=3s$, \fmref{5.3.3} yields $\mbox{dim}({\mathcal
H}_3(M=0,S=0))=1$, hence only $k=0$ appears for the ground state
and $\mu(a)=0$.  If $s$ is half integer and $a=3s-1/2$,
\fmref{5.3.3} yields $\mbox{dim}({\cal H}_3(M=1/2,S=1/2))=2$,
hence only $k=1, 2$ appear for the ground state and
$\mu(a)=1$. For all other cases, $\mu(a)>1$ and all shift
quantum numbers $k=0, 1, 2$ occur.  This completes the proof of
\fmref{5.3.0}.

\subsection{$a=2$ and odd $N$}
\label{sec-4-4}

In this subsection all states considered will be in the subspace
${\mathcal H}(M=Ns-2)$, $N$ being odd. We will prove a weaker
statement than k-rule~\ref{c-1}, namely
\begin{k-rule}
\label{5.4.1}
If there are relative ground states of $\op{H}$ with $k\neq 0$
  then there are exactly two such states with $k=1$ and $k=-1$.
\end{k-rule}
We think that the possibility $k=0$ can be excluded for $N>3$,
but the proof of this apparently requires a more
detailed analysis of the energy spectrum and will be published
elsewhere.  The situation in the case $a=2$ is greatly
simplified due to the following fact
\begin{eqnarray}\label{5.4.2}
\op{T}\ket{\psi} = \ket{\psi}
\quad\Rightarrow\quad   \op{G}\ket{\psi} =\op{G}^\dagger
\ket{\psi}
\ .
\end{eqnarray}
To prove this we define the unitary reflection
operator $\op{R}$ by linear extension of
\begin{eqnarray}\label{5.4.3}
\op{R}
\ket{m_1,m_2,\ldots,m_N}
\equiv
\ket{m_N,m_{N-1},\ldots,m_1}
\ .
\end{eqnarray}
Obviously,
\begin{eqnarray}\label{5.4.4}
\op{R}\op{G}\op{R}=\op{G}^\dagger
\ .
\end{eqnarray}
For $a=2$ any reflected product state can also be obtained by a
suitable shift, i.~e.~
\begin{eqnarray}\label{5.4.5}
\op{R} \ket{\vec{m}}
=
\op{T}^{n(\vec{m})} \ket{\vec{m}}
\ .
\end{eqnarray}
Hence $\op{R}$ maps any cycle $\{\ket{\vec{m}},
\op{T}\ket{\vec{m}}, \ldots, \op{T}^{N-1}\ket{\vec{m}}\}$ onto
itself and thus leaves states $\ket{\psi}$ with
$\op{T}\ket{\psi} = \ket{\psi}$, i.~e.~with shift quantum number
$k=0$, invariant.  Now assume $\op{T}\ket{\psi} =
\ket{\psi}$. We conclude $\op{G}^\dagger \ket{\psi} =
\op{R}\op{G}\op{R} \ket{\psi} = \op{R}\op{G} \ket{\psi} =
\op{G}\ket{\psi}$, since $\op{T}\op{G} \ket{\psi} = \op{G}\op{T}
\ket{\psi} =\op{G} \ket{\psi}$. This concludes the proof of
\fmref{5.4.2}.

In the following $E_{\text{FP}}(x)$ denotes the lowest
eigenvalue of the Frobenius-Perron Hamiltonian
$\op{H}_{\text{FP}}(x)$ as defined by Eq.~\ref{E-4-4}. Since
$[\op{H}_{\text{FP}}(x),\op{T}]=0$ there exists a complete
system of common eigenvectors of $\op{H}_{\text{FP}}(x)$ and
$\op{T}$. Especially, for $x<0$ the eigenvector corresponding to
$E_{\text{FP}}(x)$ will have positive components in the product
basis \fmref{E-2-2} and hence the shift quantum number $k=0$.

By using arguments based on the Ritz variational principle one
shows easily
\begin{eqnarray}\label{5.4.8}
x<y<0
\quad\Rightarrow\quad
E_{\text{FP}}(x)<E_{\text{FP}}(y)
\ ,
\end{eqnarray}
and
\begin{eqnarray}\label{5.4.9}
x\neq 0
\quad\Rightarrow\quad
E_{\text{FP}}(-|x|)<E_{\text{FP}}(|x|)
\ .
\end{eqnarray}
Equivalent to k-rule~\ref{5.4.1} is the corresponding statement
on $\op{H}_{\text{B}}$: If there are relative ground states of
$\op{H}_{\text{B}}$ with $k_{\text{B}}\neq 1$, then there are
exactly two such states with $k_{\text{B}}=0$ and
$k_{\text{B}}=2$.

Note that in our case $k_{\text{B}}= k+2 \frac{N+1}{2} = k+1
\text{ mod }N$.  Due to (\ref{E-4-7}) and (\ref{5.4.2})
$\op{H}_{\text{B}}$ equals $\op{H}_{\text{FP}}(-\cos
\frac{\pi}{N})$, if restricted to the sector $k=0$. The ground
state in this sector is non-degenerate according to the theorem
of Frobenius-Perron and will be denoted by $\ket{\Phi}$. It
remains to show that
\begin{description}
\item[(A)] $\ket{\Phi}$ is also a ground state of
$\op{H}_{\text{B}}$ in the whole subspace
$\left\{k_{\text{B}}=1\right\}^\perp$ which is orthogonal to the
$k_{\text{B}}=1$ sector, and
\item[(B)] any other relative ground state of
$\op{H}_{\text{B}}$ has $k_{\text{B}}=1$ or $k_{\text{B}}=2$.
\end{description}

The relative ground state of $\op{H}_{\text{B}}$ with
$k_{\text{B}}=2$ will then be non-degenerate too. This is easily
proven by re-translating into the $\op{H}$-picture and employing
the $+k \leftrightarrow -k$ symmetry.

In order to prove (A) we consider an arbitrary eigenvalue $E$ of
$\op{H}_{\text{B}}$ in ${\mathcal H}(M=Ns-2)$ which has not the
shift quantum number $k_{\text{B}}=1$. We have to show that
\begin{eqnarray}\label{5.4.9a}
E\geq E_{\text{FP}}(-\cos \frac{\pi}{N})
\ .
\end{eqnarray}
$E$ is also an eigenvalue of $\op{H}$ corresponding to an
eigenvector $\ket{\psi}$ with shift quantum number $k\neq 0$.
Since $N$ is odd, there exists an integer $\ell\ne 0$, unique
modulo $N$, such that $2\ell=N-k \mod N$. According to
\fmref{E-4-2-D}, $\ket{\phi}\equiv \op{U}^\ell \ket{\psi}$
satisfies
\begin{eqnarray}\label{5.4.10}
\op{T} \ket{\phi} = \ket{\phi},
\end{eqnarray}
and, using \fmref{E-4-3} together with \fmref{5.4.2},
\begin{eqnarray}\label{5.4.11}
\op{U}^\ell \op{H} \op{U}^{\dagger\ell} \ket{\phi}
=
E\ket{\phi}
=
\op{H}_{\text{FP}}(\cos \alpha\ell)\ket{\phi}
\ ,
\end{eqnarray}
where $\alpha\equiv2\pi/N$. Hence
\begin{eqnarray}\label{5.4.12}
E
\ge
E_{\text{FP}}(\cos \alpha\ell)
\ ,
\end{eqnarray}
by the definition of $E_{\text{FP}}(x)$.  If $ \cos \alpha\ell >0$,
\fmref{5.4.8} and \fmref{5.4.9} yield
\begin{eqnarray}\label{5.4.13}
E_{\text{FP}}(\cos \alpha\ell)
&\ge& E_{\text{FP}}(-\cos \alpha\ell)
\\
&&=
E_{\text{FP}}(\cos(\pi- \alpha\ell))
\ge E_{\text{FP}}(-\cos \frac{\pi}{N})
\ ,
\nonumber
\end{eqnarray}
since $\ell\ne 0$. For $\cos \alpha\ell <0$ the analogous
inequality follows directly from \fmref{5.4.8}. Hence
\begin{eqnarray}\label{5.4.14}
E\ge E_{\text{FP}}(-\cos \frac{\pi}{N})
\ ,
\end{eqnarray}
and the proof of (A) is complete.

Turning to the proof of (B) we note that, because of the strict
inequalities (\ref{5.4.8}) and (\ref{5.4.9}),
$E=E_{\text{FP}}\left(\cos
\alpha\ell\right)=E_{\text{FP}}\left(-\cos \frac{\pi}{N}\right)$
is only possible if
\begin{eqnarray}\label{5.4.15}
\cos \frac{2\pi\ell}{N}
&=&
\cos\alpha \ell = - \cos \frac{\pi}{N}
\ .
\end{eqnarray}
Using $2\ell = N-k \text{ mod } N$, after some elementary
calculations this can be shown to be equivalent to
\begin{eqnarray}\label{5.4.16}
k
&=&
\pm 1\text{ mod } N
\ ,
\end{eqnarray}
i.~e.~
\begin{eqnarray}\label{5.4.17}
k_{\text{B}}
&=&
0,2  \text{ mod } N
\ ,
\end{eqnarray}
which completes the proof of (B) and k-rule~\ref{5.4.1}.

\subsection{Haldane systems}
\label{sec-4-5}

One idea to prove part of the k-rule~\ref{c-2} for odd $N$
would be to show that one of the relative ground states has an
overlap with another eigenstate of the shift operator whose
shift quantum number is known to be zero.  A good candidate
would be the relative ground state of
$\op{H}_{\text{FP}}(-\cos\pi/N)$ \fmref{E-4-4} in ${\mathcal
H}(M)$ which has $k=0$. If this state has overlap with a
relative ground state of $\op{H}_{\text{B}}$ \fmref{E-4-7} the
latter also possesses $k=0$.

Let $\op{V}=\op{U}^{(N+1)/2}$, $\ket{\Psi_0}$ and
$\ket{\hat{\Psi}_0}=\op{V}\ket{\Psi_0}$ be one of the
relative ground states of $\op{H}$ \fmref{E-2-1} and
$\op{H}_{\text{B}}$ \fmref{E-4-3},
respectively. $\ket{\Psi_{\text{FP}}}$ denotes the relative
ground state of $\op{H}_{\text{FP}}$.  Then part of the k-rule is
implied by the following
\begin{k-rule}
\label{c-3}
$\ket{\Psi_{\text{FP}}}$ has a non-vanishing
$\op{H}_{\text{B}}$-ground-state component, i.~e.~
$\braket{\Psi_{\text{FP}}}{\hat{\Psi}_0}\ne 0$.
\end{k-rule}
The validity of this k-rule would immediately follow from
the sufficient (but not necessary) inequality
\begin{eqnarray}
\label{E-4-12}
E_{\text{FP}}-E_0
&<&
E_1 - E_0
\ ,
\end{eqnarray}
where $E_1$ is the energy of the first excited state above the
relative ground state in ${\mathcal H}(M)$ and
\begin{eqnarray}
\label{E-4-13}
E_{\text{FP}}
=
\bra{\Psi_{\text{FP}}}
\op{H}_{\text{B}}
\ket{\Psi_{\text{FP}}}
=
\bra{\Psi_{\text{FP}}}
\op{H}_{\text{FP}}
\ket{\Psi_{\text{FP}}}
\ .
\end{eqnarray}
As a substitute for the lacking proof of k-rule~\ref{c-3} we
submit the inequality \fmref{E-4-12} to some numerical tests,
see section~\ref{sec-5}.

Looking at the large $N$ behavior it is nevertheless possible to
devise an asymptotic proof for systems which possess a finite
energy gap in the thermodynamic limit $N\rightarrow\infty$.
These systems are called ``Haldane systems".  According to
Haldane's conjecture\cite{Hal:PL83,Hal:PRL83} spin rings with an
integer spin quantum number $s$ possess such gaps.

To start with the proof, let us look for an upper bound to
$E_{\text{FP}}-E_0$. Take $\ket{\Psi_0}$ to be a ground state of
$\op{H}$ with real coefficients with respect to the product
basis $\{\ket{\vec{m}}\}$. Evidently,
\begin{eqnarray}
\label{E-3-14}
E_{\text{FP}}
&\le&
\bra{\Psi_0}\op{V}^\dagger\op{H}_{\text{FP}}\op{V}\ket{\Psi_0}
\\
&\le&
E_0
+
i \sin\left( \frac{\pi}{N} \right)
\nonumber
\\
&&\quad\times
\bra{\Psi_0}\op{V}^\dagger
\left\{ \op{G} - \op{G}\right\}
\op{V}^\dagger\ket{\Psi_0}
\nonumber
\ .
\end{eqnarray}
Further, in view of \fmref{E-4-2-E}
\begin{eqnarray}
\label{E-3-15}
\op{V}^\dagger\left\{ \op{G} - \op{G}^\dagger\right\}\op{V}
&=&
-\left\{\text{e}^{i\frac{\pi}{N}} \op{G}
-
\text{e}^{-i\frac{\pi}{N}}\op{G}^\dagger\right\}
\ ,
\end{eqnarray}
and, because $\braket{\vek{m}}{\Psi_0}$ being real,
$\bra{\Psi_0}\op{G} -
\op{G}^\dagger\ket{\Psi_0}=0$. Therefore,
\begin{eqnarray}
\label{E-3-16}
E_{\text{FP}}-E_0
&\le&
\sin^2\left( \frac{\pi}{N} \right)
\bra{\Psi_0}\op{G} + \op{G}^\dagger\ket{\Psi_0}
\ .
\end{eqnarray}
A rough upper estimate for the operator norm of $\{\op{G} +
\op{G}^\dagger\}$ in ${\mathcal H}(M=Ns-a)$ can be deduced from
the well-known Ger\v{s}gorin bounds for matrix eigenvalues
(c.~f.~\cite{Lan:69}, 7.2):
\begin{eqnarray}
\label{E-3-17}
||\op{G} + \op{G}^\dagger||
\le
2\, f(s)\,\min(a, N, 2Ns-a)
\ ,
\end{eqnarray}
where
\begin{eqnarray}
\label{E-3-18}
f(s)
&=&
\begin{cases}
(s+\frac{1}{2})^2               & s\ \text{half integer}\\
(s+\frac{1}{2})^2 - \frac{1}{4} & s\ \text{integer}
\end{cases}
\ .
\end{eqnarray}
We therefore conclude
\begin{eqnarray}
\label{E-3-19}
E_{\text{FP}}-E_0
&\le&
2\,N\,
\sin^2\left( \frac{\pi}{N} \right)
f(s)
\ .
\end{eqnarray}
Thus with increasing $N$ $(E_{\text{FP}}-E_0)$ approaches zero
at least like $1/N$ and therefore, above some $N_0$,
$(E_{\text{FP}}-E_0)$ must be smaller than the Haldane gap
$(E_1-E_0)$.

One would of course like to accomplish a similar proof for half
integer spin systems, but in this case $(E_1-E_0)$ drops like
$1/N$ itself as given by the Wess-Zumino-Witten model, see
e.g. Ref.~\onlinecite{AGS:JPA89}. Thus for such systems a
careful analysis of the coefficient in front of the $1/N$ might
be very valuable. As shown in the next section, numerical
investigations indicate that $(E_{\text{FP}}-E_0)$
approaches zero faster than $(E_1-E_0)$.

\section{Numerical studies}
\label{sec-5}

\begin{table}[!ht]
\begin{center}
\vspace*{1mm}
\begin{tabular}{|cc||r|r|r|r|l|}
\hline
&$s$&\multicolumn{4}{c|}{$N$}&\\
&& 3&5&7&9&\\
\hline\hline
\multirow{36}{0mm}{}
 &              & -1.5 & -3.736 & -5.710 & -7.595 & $E_0/|J|$\\
 & $\frac{1}{2}$& -1.5 & -3.736 & -5.706 & -7.589 & $E_{\text{FP}}/|J|$\\
 &              &  1.5 & -1.5   & -3.612 & -5.872 & $E_1/|J|$\\
\cline{2-7}
 &              & -6.0   & -13.062 & -19.144 & -24.960 & $E_0/|J|$\\
 & $      1    $& -5.162 & -12.180 & -18.338 & -24.235 & $E_{\text{FP}}/|J|$\\
 &              & -4.0   & -11.133 & -17.431 & -23.420 & $E_1/|J|$\\
\cline{2-7}
 &              & -10.5  & -24.865 & -37.370 & -49.296 & $E_0/|J|$\\
 & $\frac{3}{2}$& -9.788 & -24.095 & -36.663 & -48.658 & $E_{\text{FP}}/|J|$\\
 &              & -7.5   & -22.237 & -35.199 & -47.458 & $E_1/|J|$\\
\cline{2-7}
 &              & -18.0   & -42.278 & -63.315 & $-83.364^{\dagger}$ & $E_0/|J|$\\
 & $      2    $& -16.506 & -40.615 & -61.789 & $-81.989^{\dagger}$ & $E_{\text{FP}}/|J|$\\
 &              & -16.0   & -40.356 & -61.663 & $-81.934^{\dagger}$ & $E_1/|J|$\\
\cline{2-7}
 &              & -25.5   & -62.168 & -94.160 & $-124.63^{\dagger}$ & $E_0/|J|$\\
 & $\frac{5}{2}$& -24.188 & -60.699 & -92.814 & $-123.42^{\dagger}$ & $E_{\text{FP}}/|J|$\\
 &              & -22.5   & -59.538 & -92.006 & $-122.83^{\dagger}$ & $E_1/|J|$\\
\cline{2-7}
 &              & -36.0   & -87.666 & $-132.68^{\dagger}$ & $-175.55^{\dagger}$ & $E_0/|J|$\\
 & $      3    $& -33.936 & -85.325 & $-130.55^{\dagger}$ & $-174.18^{\dagger}$ & $E_{\text{FP}}/|J|$\\
 &              & -34.0   & -85.747 & $-131.06^{\dagger}$ & $-173.66^{\dagger}$ & $E_1/|J|$\\
\hline
\end{tabular}
\vspace*{5mm}
\end{center}
\caption[]{Lowest energy eigenvalues of the Heisenberg
  Hamiltonian ($E_0$,$E_1$) as well as of the respective Frobenius-Perron
  Hamiltonian ($E_{\text{FP}}$) for various odd $N$ and $s$;
  $\dagger$ -- projection method.\cite{Man:RMP91}
  Note that we find $E_0 \le E_{\text{FP}}< E_1$ for all $N$ if $s\le 5/2$.
}\label{T-2}
\end{table}

\begin{figure}[!ht]
\begin{center}
\vspace*{1mm}
\epsfig{file=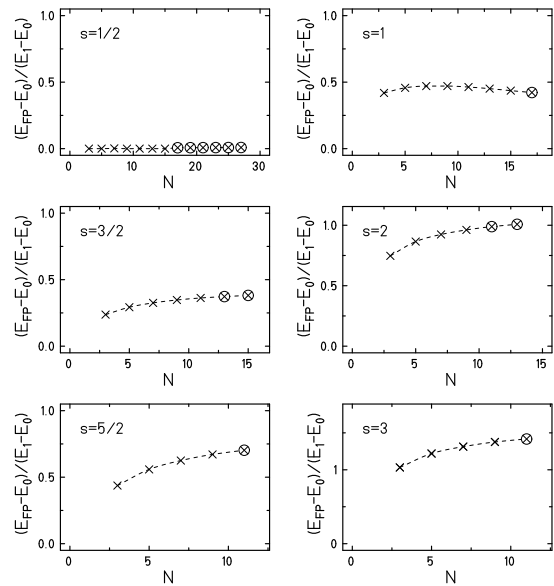,width=75mm}
\vspace*{1mm}
\caption[]{Dependence of $(E_{\text{FP}}-E_0)/(E_1-E_0)$ on $N$
  for various $s$. Crosses denote values obtained by exact
  diagonalization or projection method, circled crosses denote
  values obtained by a L\'anczos method. For $s=1/2$, where
  $[\op{G},\op{G}^\dagger]=0$, the ratio
  $(E_{\text{FP}}-E_0)/(E_1-E_0)$ is extremely small, i.~e. $\approx 
  10^{-2}$.}
\label{F-1}
\end{center}
\end{figure}

The question \fmref{E-4-12} whether $(E_{\text{FP}}-E_0)<(E_1 -
E_0)$ holds in ${\mathcal H}(M$ with minimal $|M|$ was
investigated numerically. For some of the investigated rings the
respective energies are given in Table~\ref{T-2}.

Figure \ref{F-1} shows the ratio $(E_{\text{FP}}-E_0)/(E_1-E_0)$
for rings with $s=1/2,\dots,3$ and various $N$. This ratio is
smaller than one for $s=1/2,1,3/2,5/2$ for all investigated
$N$. Only for $s=2,3$ the ratio reaches values above
one. Nevertheless, as discussed in the previous section, in the
cases of integer $s$ this ratio must approach zero like $1/N$
since $(E_1-E_0)$ approaches the Haldane gap. But also in the
cases of half integer spin one is led to anticipate that the
curves rising with $N$ for small $N$ will bend down later and
approach zero for large $N$. DMRG calculations could help to
clarify this question.

\section{Generalization to other spin models}
\label{sec-6}

It is a legitimate question whether the k-rule holds for
Heisenberg spin rings only or whether it is valid for a broader
class of spin Hamiltonians. In order to clarify this question we
investigate the following XXZ-Hamiltonian
\begin{eqnarray}
\label{E-6-1}
\op{H}(\delta)
&=&
\delta\cdot\op{\Delta} + \op{G} + \op{G}^\dagger
\ ,
\end{eqnarray}
for various values of $\delta$. The case $\delta=1$ corresponds
to the original Heisenberg Hamiltonian \fmref{E-2-5},
$\delta\rightarrow\infty$ results in the antiferromagnetic Ising
model, $\delta\rightarrow -\infty$ in the ferromagnetic Ising
model, and $\delta=0$ describes the XY-model.

We have numerically investigated the cases of
$\delta=-1000,-1,0, 0.5,1000$ for $s=1/2,\dots, 5/2$ and
$N=5,\dots,8$. For $|\delta|\le 1$ no violation of the k-rule
was found, whereas the k-rule is violated for $\delta=\pm
1000$.

\begin{figure}[!ht]
\begin{center}
\epsfig{file=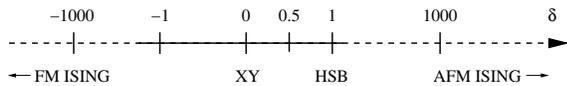,width=75mm}
\vspace*{1mm}
\caption[]{Solid line: Estimated validity of the k-rule for various
  parameters $\delta$ of the Hamiltonian \fmref{E-6-1}. The
  numbers denote the cases which have been examined
  numerically. The k-rule is violated for $\delta=\pm 1000$,
  no violation was found for $|\delta|\le 1$.}
\label{F-2}
\end{center}
\end{figure}

In the limiting case of the Ising model the k-rule \ref{c-1}
is in general violated. Any product state $\ket{\vec{m}}$ will
be an eigenstate of the Ising Hamiltonian and the shifted states
$\op{T}^\nu\ket{\vec{m}}$ belong to the same eigenvalue
$E_{\vec{m}}$. The set of the corresponding shift quantum
numbers then depends on the degree of symmetry of
$\ket{\vec{m}}$: Let $n$ denote the smallest positive integer
such that $\op{T}^n\ket{\vec{m}}=\ket{\vec{m}}$.  Clearly, $n$
divides $N$.  Then the corresponding shift quantum numbers will
be of the form $k=\frac{N}{n}\ell \text{ mod } N,\; \ell =
0,1,2,\ldots$.  In most cases, $n=N$ and hence all possible
shift quantum numbers will occur, which violates \ref{c-1}. On
the other hand consider the total ground state
$\ket{\uparrow,\downarrow,\uparrow,\downarrow,\ldots}$ of an
even $s=1/2$ antiferromagnetic Ising spin ring.  Here we have
$n=2$ and only the shift quantum numbers $k=0, \frac{N}{2}$
occur, also contrary to \xref{E-1-1}. Figure~\ref{F-2}
summarizes our findings as a graphics.

It is not clear at which $\delta$ exactly the k-rule breaks
down, this quantum phase transition might very much depend on
$N$ and $s$. It is then an open question whether another
k-rule takes over.

Finally we would like to mention that the exactly solvable
$s=1/2$ $XY$-model
\cite{LSM:AP61,Mat88}
satisfies the k-rule \fmref{E-1-1}. This model is
essentially equivalent to a system of $a$ non-interacting
Fermions. More precisely, for odd $a$ its energy eigenvalues
are of the form
\begin{eqnarray}
\label{E-6-2}
E_{\vec{k}}^{\text{(odd)}}
&=&
2
\sum_{\nu=1}^a \cos\left(\frac{2\pi}{N} k_\nu\right)
\ ,\ k_\nu \text{ integer}
\ ,
\end{eqnarray}
with corresponding shift quantum numbers
\begin{eqnarray}
\label{E-6-3}
k
&=&
\sum_{\nu=1}^a  k_\nu
\text{ mod } N
\ .
\end{eqnarray}
Relative ground state configurations $\vec{k}$ for
$a=1,3,5,\ldots$  and odd $N$ are, for example,
\begin{eqnarray}
\label{E-6-4}
\vec{k}
&=&
\left(\frac{N+1}{2} \right),
\left(\frac{N\pm 1}{2},\frac{N+3}{2} \right),
\\
&&
\left(\frac{N\pm 1}{2},\frac{N\pm 3}{2},\frac{N+5}{2} \right)
\ , \ldots
\nonumber
\end{eqnarray}
This leads to the shift quantum numbers
\begin{eqnarray}
\label{E-6-5}
k
&=&
\frac{N+1}{2}, \frac{N+3}{2}, \frac{N+5}{2} 
\ , \ldots
\end{eqnarray}
in accordance with  \fmref{E-1-1}. Similarly, the values
\begin{eqnarray}
\label{E-6-6}
k
&=&
\frac{N-1}{2}, \frac{N-3}{2}, \frac{N-5}{2} 
\ ,\ldots
\end{eqnarray}
are realized. In the case of even $a$ we have
\begin{eqnarray}
\label{E-6-7}
E_{\vec{k}}^{\text{(even)}}
&=&
2
\sum_{\nu=1}^a \cos\left(\frac{2\pi}{N} \frac{2k_\nu+1}{2}\right)
\ ,\ k_\nu \text{ integer}
\ ,
\end{eqnarray}
with corresponding shift quantum numbers
\begin{eqnarray}
\label{E-6-8}
k
&=&
\sum_{\nu=1}^a  \left(k_\nu+\frac{1}{2}\right)
\text{ mod } N
\ ,
\end{eqnarray}
and the k-rule \ref{c-1} follows analogously.

\section*{Acknowledgments}

We thank Shannon Star for motivating discussions and Ian Affleck
for pointing out some useful literature.


\end{document}